\documentclass[pre,preprint,showpacs,aps]{revtex4}
\usepackage{amsmath,amssymb,graphicx}
\begin{document} 

\title[An unusual Li\'enard type nonlinear oscillator with properties ...]
{An unusual Li\'enard type nonlinear oscillator with properties of a linear
harmonic oscillator}

\author{V. K. Chandrasekar, M. Senthilvelan and M. Lakshmanan}

\affiliation{Centre for Nonlinear Dynamics, Department of Physics, 
Bharathidasan University, Tiruchirappalli - 620 024, India}

\date{\today} 

\begin{abstract}
A Li\'enard type nonlinear oscillator of the form
$\ddot{x}+kx\dot{x}+\frac{k^2}{9}x^3+\lambda_1 x=0$,
which may also be considered as a generalized Emden type equation, is shown to
possess unusual nonlinear dynamical properties. It is shown to admit explicit 
nonisolated periodic orbits of conservative Hamiltonian type for $\lambda_1>0$. 
These periodic orbits
exhibit the unexpected property that the frequency of oscillations is
completely independent of amplitude and continues to remain as that of the 
linear harmonic oscillator. This is completely contrary to the standard 
characteristic property of nonlinear oscillators. Interestingly, the system 
though appears deceptively a dissipative type for $\lambda_1\leq0$ does admit a
conserved Hamiltonian description, where the characteristic decay time is also
independent of the amplitude. 
The results also show that the criterion for conservative Hamiltonian system 
 in terms of divergence of flow function needs to be generalized. 
 
\end{abstract}

\maketitle
\section{Introduction}
\label{1}

Nonlinear oscillator systems are ubiquitous and they model numerous 
physical phenomena 
ranging from  atmospheric physics, condensed matter, nonlinear optics to 
electronics, plasma physics, biophysics, evolutionary biology, 
etc. \cite{ Nay:1989, Guc:1983, Tab:1989, Wig:2003, Lak:2003}. One of the 
most important characteristics of nonlinear
oscillations is the amplitude or initial condition dependence of frequency for
nonisolated periodic orbits \cite{Nay:1989, Tab:1989, Lak:2003}. For example, 
for the cubic anharmonic oscillator 
\begin{eqnarray}
\ddot{x}+\omega_0^2x+\beta x^3=0, \qquad (\omega_0^2,\beta >0), 
\label {lam100}
\end{eqnarray}
where dot denotes differentiation with respect to $t$, the general (periodic) 
solution is $x(t)=A\;cn(\Omega t+\delta)$, where $A$ and $\delta$ are arbitrary 
constants with the modulus squared of the Jacobian elliptic function 
$m^2=\frac{\beta A^2}{2(\omega_0^2+\beta A^2)}$ and frequency 
$\Omega=\sqrt{\omega_0^2+\beta A^2}$. 
In the case of
limit cycle oscillations the dependence of the initial condition is manifested
in the
form of suitable transient time to reach the asymptotic state. For chaotic 
oscillations,
of course, there is the sensitive dependence on initial conditions. In this 
paper we
identify a physically interesting and simple nonlinear oscillator of the Li\'enard
type, which is also a generalized Emden type equation, that admits for a
particular sign of the control parameter 
nonisolated conservative periodic oscillations, exhibiting the remarkable fact 
that the
frequency of oscillation is completely independent of amplitude and remains the
same as that of the linear harmonic oscillator, thereby
showing
that the amplitude dependence of frequency is not necessarily a fundamental
property of nonlinear dynamical phenomena. We also show that rewriting the 
underlying
equation  of motion as a system of two first order coupled nonlinear
differential equations the basic
criterion for conservative system in terms of the divergence of the flow 
function has to
be generalized.

 We consider a nonlinear oscillator system of the form 
\begin{eqnarray}            
\ddot{x}+kx\dot{x}+\frac {{k}^2}{9}x^3+\lambda_1 x=0,\label {lam101}
\end{eqnarray}
which is of the Li\'enard type  $\ddot{x}+f(x)\dot{x}+g(x)=0$, where $f(x)=kx$
and $g(x)=\frac {{k}^2}{9}x^3+\lambda_1 x$ in the present case. Eq.~(\ref{lam101})
can also be interpreted as the cubic anharmonic oscillator defined by 
Eq.~(\ref{lam100}) (with $\omega_0^2=\lambda_1$ and $\beta=\frac{k^2}{9}$), 
but acted upon on by a strong damping type nonlinear force
$kx\dot{x}$.  This type of equation is also well
studied in the literature for almost two decades as a generalized form of  
Emden type equation occurring in the study of equilibrium configurations of a
spherical cloud acting under the mutual attraction of its molecules and subject
to the laws of thermodynamics \cite{Dix:1990}. Eq.~(\ref{lam101}) is a special 
case of a
general second order nonlinear differential equation possessing eight Lie 
point symmetries \cite{mah:1985} and that it is linearizable. In 
particular the special case $\lambda_1=0$ has been well analyzed by Mahomed 
and Leach \cite{mah:1985} and explicit forms of the eight symmetry generators 
satisfying $sl(3,R)$ algebra have been well documented.  The latter case 
$(\lambda_1=0)$ has the exact general solution
\begin{eqnarray}
x(t)=\frac{t+I_1}{\frac{k}{6}t^2+\frac{I_1k}{3}t+I_2},
\label {lam102}
\end{eqnarray}
where $I_1$ and $I_2$ are the two integrals of motion with the explicit forms
\begin{subequations}
\begin{eqnarray}
I_1=-t+\frac{x}{\frac{k}{3}x^2+\dot{x}},\;\;\\
I_2=\frac{k}{6}t^2+\frac{1-\frac{k}{3}tx}{\frac{k}{3}x^2+\dot{x}}.
\end{eqnarray}
\label {lam103}
\end{subequations}
Obviously for $\lambda_1=0$ case the initial value 
problem of Eq.~(\ref{lam101}) appears to be a dissipative
nonlinear system.

In the following we show that the nonlinear oscillator equation (\ref{lam101})
possesses several unusual and interesting features, which are contrary to the
standard characteristic features associated with usual nonlinear
oscillators. For example, in the case $\lambda_1>0$ we prove that the system 
(\ref{lam101}) possesses oscillatory motion whose frequency is completely
independent of the amplitude.  Thus we prove that the standard notion of 
{\it amplitude dependent frequency of oscillations is not a necessary condition for 
nonlinear oscillators}. Moreover, we also point out that the system
(\ref{lam101}) admits a Lagrangian formulation and from which a conservative
Hamiltonian can be obtained. Further, we also find that the standard test 
(divergence of the flow function) to identify whether the given system is 
conservative or not fails here. To overcome this situation we propose that
the necessary condition for a system to be a conserved  Hamiltonian is that the 
time average of the divergence of the flow function should vanish instead of 
the flow function itself vanishing. More interestingly, we also find that the system 
(\ref{lam101}) when $\lambda_1\leq0$ also admits a Lagrangian as well as a conserved 
Hamiltonian
description similar to the case $\lambda_1>0$, eventhough the general solution
shows a dissipative/damping/frontlike aperiodic structure, depending on the
initial condition. The divergence of the flow function for
the case $\lambda_1\leq0$ can be negative for all times in spite the conserved
Hamiltonian nature. Again to redeem the situation, one requires the time average
of the divergence of the flow function to be vanishing. Finally, we also point
out that the special features of the system can be traced to the existence of
certain linearizing and canonical and nonlocal transformations, which points 
out to generalization of the results in different directions.

The plan of the paper is as follows. In the following section, we utilize the so
called modified 
Prelle-Singer procedure and derive two functionally independent integrals of
motion for the Eq. (\ref{lam101}). In Sec.~\ref{sec3}, we show that for the case 
$\lambda_1>0$ the system (\ref{lam101}) admits periodic solutions and 
construct the associated Lagrangian as well as
Hamiltonian for this system. We discuss the unusual features associated with the
system in Sec.~\ref{sec4} and point out the amplitude independence of the 
frequency of oscillations as well as the nonzero value of the divergence of 
the flow function in spite of its conservative nature. We analyse the nature 
of solutions in the regimes $\lambda_1=0$ and $\lambda_1<0$ in Sec.~\ref{sec5} 
and point out that in spite of the decaying/damping nature or frontlike 
aperiodic nature of the solutions, 
the system admits Lagrangian and Hamiltonian descriptions. In Sec.~\ref{sec6}, 
we linearize the Eq. (\ref{lam101}) 
through different kinds of transformations (invertible point, nonlocal and
canonical transformation) to trace the unusual features of 
the system. In Sec.~\ref{sec7}, we generalize Eq. (\ref{lam101}) and compare the
dynamics with certain other interesting nonlinear oscillators. Finally, 
we present our conclusions in Sec.~\ref{sec8}.

\section{Extended Prelle-Singer procedure}
\label{sec2}
Now let us consider the general case $\lambda_1\ne0$. One can proceed to solve 
the equation explicitly  by the so called modified Prelle-Singer (P-S) method
\cite{Dua:2001,Cha:2004} which identifies the integrals of motion and explicit 
solution, if they exist. One can also use other methods as well; however we 
find that the P-S method is quite convenient \cite{Cha:2004} to
obtain both the integrals of motion and solution explicitly. Assuming the 
existence
of an integral $I=I(t,x,\dot{x})$ for Eq.~(\ref{lam101}) and rewriting the 
latter as $\ddot{x}=P(x,\dot{x})$, where 
$P=-kx\dot{x}-\frac {{k}^2}{9}x^3-\lambda_1 x$, so that 
$Pdt-d\dot{x}=0$, and introducing a null term 
$S(t,x,\dot{x})\dot{x}dt-S(t,x,\dot{x})dx$, we find that on 
the solutions, $(P+S\dot{x})dt-Sdx-d\dot{x} = 0$. Since $I$ is a constant of
motion, it follows that
\begin{eqnarray}  
dI={I_t}{dt}+{I_{x}}{dx}+{I_{\dot{x}}{d\dot{x}}}=0, 
\label{met3}  
\end{eqnarray}
so that
\begin{eqnarray} 
dI=R(P+S\dot{x})dt-RSdx-Rd\dot{x}=0, 
\label{met7}
\end{eqnarray}
where $R$ is an integrating factor. Comparing equations (\ref{met3}) 
with (\ref{met7}) we have, on the solutions, the relations 
\begin{eqnarray} 
I_{t} = R(P+\dot{x}S)  \nonumber \\
I_{x} =-RS \qquad\;\;\nonumber \\
I_{\dot{x}} = -R. \quad\qquad\nonumber  
\end{eqnarray} 
Then the compatibility conditions, 
$I_{tx}=I_{xt}$, $I_{t\dot{x}}=I_{{\dot{x}}t}$,
$I_{x{\dot{x}}}=I_{{\dot{x}}x}$, require that 
\begin{eqnarray}  
D[S] = -P_x+SP_{\dot{x}}+S^2,\nonumber\\
D[R] = -R(S+P_{\dot{x}}),\qquad\nonumber\\
R_x  = R_{\dot{x}}S+RS_{\dot{x}},\qquad\;\;
\label{met11}
\end{eqnarray}
where 
\begin{eqnarray} 
 D=\frac{\partial}{\partial{t}}+
\dot{x}\frac{\partial}{\partial{x}}+P\frac{\partial}
{\partial{\dot{x}}}. \nonumber
\end{eqnarray}

Solving (\ref{met11}) systematically for $S$ and $R$, one 
can write down the form of the integral of motion,
\begin{eqnarray}
 I= r_1
  -r_2 -\int \left[R+\frac{d}{d\dot{x}} \left(r_1-r_2\right)\right]d\dot{x},
  \label{met13}
\end{eqnarray}
where 
\begin{eqnarray} 
r_1 = \int R(P+\dot{x}S)dt,\quad
r_2 =\int (RS+\frac{d}{dx}r_1) dx, \nonumber
\end{eqnarray}
for the given form $P=-kx\dot{x}-\frac{{k}^2}{9}x^3-\lambda_1 x$. We find two
independent sets of compatible solutions of Eq.~(\ref{met11}), namely,
\begin{subequations}
\begin{align}
&S_1 =\frac{-\dot{x}+\frac{k}{3}x^2}{x},\quad
R_1 = \frac{xe^{-2\sqrt{-\lambda_1}t}}{(\dot{x}+\frac{k}{3}x^2-
\sqrt{\lambda_1}x)^2}, \\
&S_2 =\frac{(\frac{k}{3}x+\sqrt{-\lambda_1})^2-\frac{k}{3}\dot{x}}
{\frac{k}{3}x+\sqrt{-\lambda_1}},\;
R_2 = \frac{(\frac{k}{3}x+\sqrt{-\lambda_1})e^{\sqrt{-\lambda_1}t}}
{(\dot{x}+\frac{k}{3}x^2+\sqrt{-\lambda_1}x)^2}.
\label{lam107}
\end{align}
\end{subequations}
Consequently, substituting $R$ and $S$ in (\ref{met13}), we find the two 
(time dependent) integrals of motions for Eq.~(\ref{lam101}) as 
\begin{subequations}
\begin{eqnarray}
I_1=e^{-2\sqrt{-\lambda_1}t}{\bigg(\frac{\dot{x}+\frac{k}{3}x^2+\sqrt{-\lambda_1}x}
{\dot{x}+\frac{k}{3}x^2-\sqrt{-\lambda_1}x}}\bigg), \\
I_2=-\frac{6}{k}e^{\sqrt{-\lambda_1}t}
{\bigg(\frac{\lambda_1+\frac{k}{3}\dot{x}+\frac{k^2}{9}x^2}
{\dot{x}+\frac{k}{3}x^2+\sqrt{-\lambda_1}x}}\bigg).
\end{eqnarray}
\label{lam117}
\end{subequations}

\section{Periodic solutions, Lagrangian and Hamiltonian Description for the 
case $\lambda_1>0$}
\label{sec3}
As we mentioned in the introduction Eq.~(\ref{lam101}) admits two different kinds of
dynamics depending on the sign of the parameter $\lambda_1$. In this section we
discuss the case $\lambda_1>0$.

\subsection{Periodic solutions}
Now we note that for $\lambda_1>0$ both the integrals $I_1$ and $I_2$ become 
complex.  To identify two real integrals, we consider the combinations
\begin{eqnarray}
\mathfrak{I}_1=\frac{4}{k^2I_1I_2^2}=
\frac{(3\dot{x}+kx^2)^2+9\omega^2 x^2}
{(3k\dot{x}+k^2x^2+9\omega^2)^2},
\label{lam118}
\end{eqnarray}
where $\omega=\sqrt{\lambda_1}$ and
\begin{eqnarray}
\mathfrak{I}_2=-\frac{2e^{i\delta}}{k|I_1I_2|}=
e^{i(\omega t+\delta)}\bigg(\frac{3\dot{x}+kx^2-3i\omega x}
{3k\dot{x}+k^2x^2+9\omega^2}\bigg),
\label{lam119}
\end{eqnarray} 
so that $\mathfrak{I}_1$ and $|\mathfrak{I}_2|$ can be taken as the two real 
integrals of Eq.~(\ref{lam101}) for $\lambda_1>0$. In Eq.~(\ref{lam119}),
$\delta$ is a phase constant.

Thus for the case $\lambda_1>0$, the integrals (\ref{lam118}) and 
(\ref{lam119}) lead to the explicit sinusoidal periodic solution (Fig. 1),
\begin{eqnarray}
x(t)=\frac{A\sin{(\omega t+\delta)}}
{1-(\frac{k}{3\omega})A\cos{(\omega t+\delta)}},\qquad\qquad\qquad \nonumber\\ 
\qquad\qquad\qquad\qquad 0\leq A <\frac{3\omega}{k},
\qquad \omega=\sqrt{\lambda_1},
\label{lam120}
\end{eqnarray}
where $A=3\omega\sqrt{\mathfrak{I}_1}$ and $\delta$ is an arbitrary constant. 
The form of the first integral
(\ref{lam118}) also establishes that the system is of conservative type. Note
that in (\ref{lam120})
\begin{eqnarray}
 x_{max}=B=\frac{A}{(1-\frac{k^2}{9\omega^2}A^2)^
{\frac{1}{2}}}\nonumber
\end{eqnarray}
and
\begin{eqnarray}
x_{min}=-\frac{A}{(1-\frac{k^2}{9\omega^2}A^2)^{\frac{1}{2}}},
\nonumber
\end{eqnarray}
 so that 
\begin{eqnarray}
A=\pm\frac{B}{(1+\frac{k^2}{9\omega^2}B^2)^{\frac{1}{2}}}\nonumber
\end{eqnarray}
and $B$ may be called 
as the amplitude of oscillations in the present case.  Also for 
$A\geq\frac{3\omega}{k}$, the solution (\ref{lam120}) becomes singular at 
finite times.

\begin{figure}[!ht]
\begin{center}
\includegraphics[width=\linewidth]{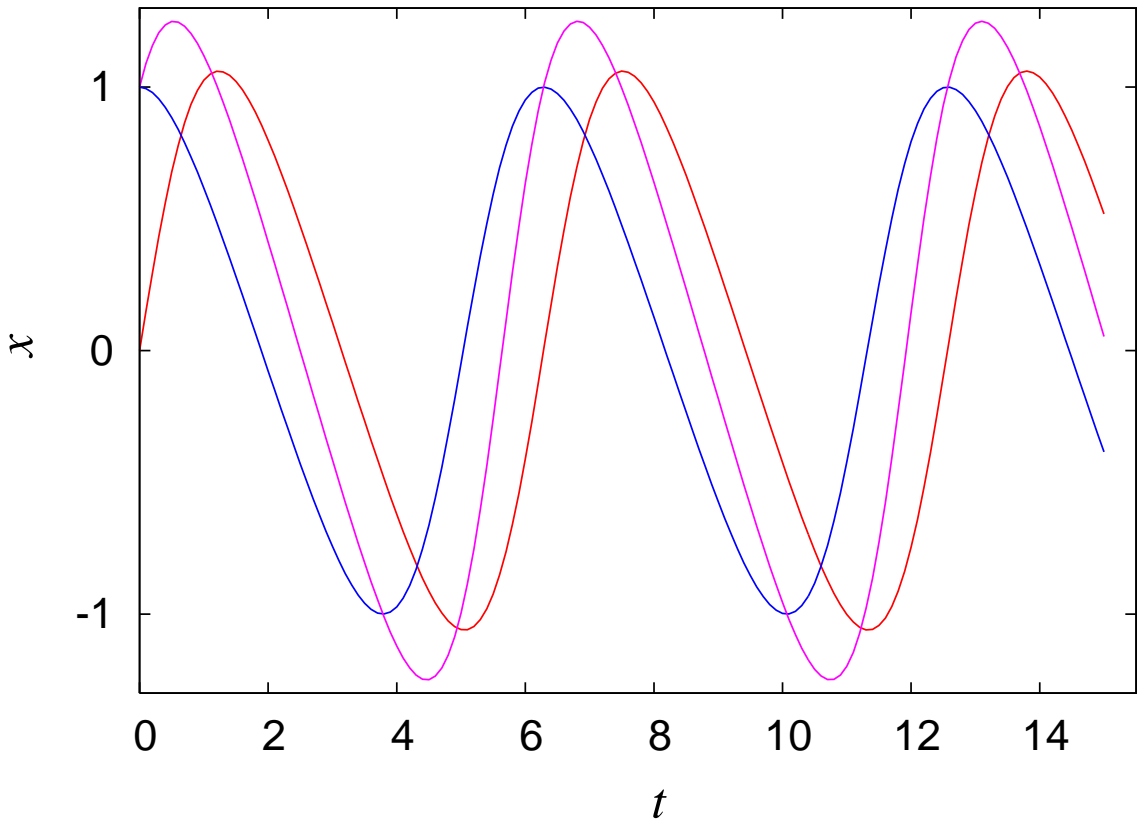}
\caption{(Color online) Harmonic periodic solutions of the nonlinear Li\'enard type oscillator
for different initial conditions 
(Eq.~(\ref{lam101})) with frequency same as that of the linear harmonic 
oscillator.}
\end{center}
\end{figure}
\begin{figure}[!ht]
\begin{center}
\includegraphics[width=\linewidth]{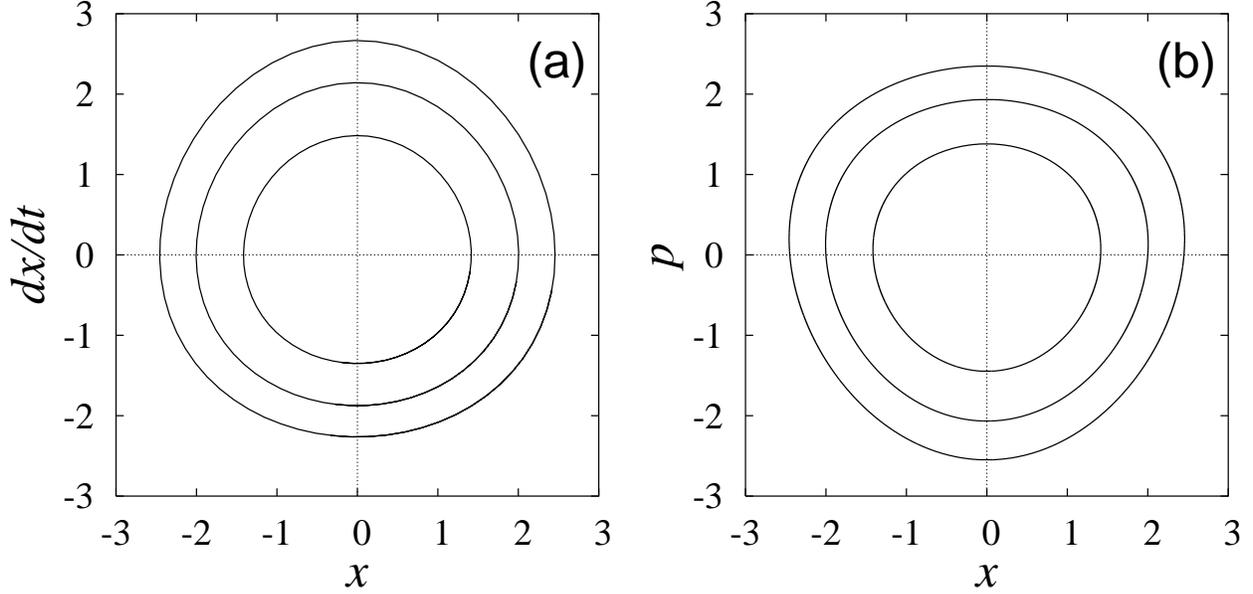}
\caption{Phase trajectories corresponding to Eq.~(\ref{lam101}) for 
$|x|<\frac{3\omega}{k}$: (a) $(x-\dot{x})$ plane, (b) $(x-p)$ plane.}
\end{center}
\end{figure}

\subsection{Lagrangian and Hamiltonian description}
The form of the time independent first integral $\mathfrak{I}_1$ as given by 
Eq.~(\ref{lam118}) suggests that one can indeed give a Lagrangian and so a Hamiltonian 
description for Eq.~(\ref{lam101}) when $\lambda_1>0$. Indeed we identify a
suitable Lagrangian for the system (\ref{lam101}) as
\begin{eqnarray} 
L=\frac{27\lambda_1^3}{2k^2}
\bigg(\frac{1}{k\dot{x}+\frac{k^2}{3}x^2+3\lambda_1}\bigg)
+\frac{3\lambda_1}{2k}\dot{x}-\frac{9\lambda_1^2}{2k^2}.
\label{lam121}
\end{eqnarray}
Then the canonically conjugate momentum can be defined as
\begin{eqnarray} 
p=\frac{ \partial L}{\partial \dot{x}}=-\frac{27\lambda_1^3}{2k}
\bigg(\frac{1}{(k\dot{x}+\frac{k^2}{3}x^2+3\lambda_1)^2}\bigg)
+\frac{3\lambda_1}{2k}.
\label{lam122}
\end{eqnarray}
Consequently, one can obtain the Hamiltonian associated with Eq.~(\ref{lam101})
as
\begin{subequations}
\begin{align} 
&H=\frac{9\lambda_1^2}{2}\bigg(\frac{(\dot{x}+\frac{k}{3}x^2)^2+\lambda_1 x^2}
{(k\dot{x}+\frac{k^2}{3}x^2+3\lambda_1)^2}\bigg)
\equiv \frac{9\lambda_1^2\mathfrak{I}_1}{2},\\
\hbox{or}\nonumber\\
&H=\frac{9\lambda_1^2}{2k^2}
\bigg(2-2(1-\frac{2kp}{3\lambda_1})^{\frac{1}{2}}
+\frac{k^2x^2}{9\lambda_1}-\frac{2kp}{3\lambda_1}-\frac{2k^3x^2p}{27\lambda_1^2}\bigg).
\end{align}
\label{lam123}
\end{subequations}
Note that in the above expressions $L,p$ and $H$, we have kept certain
prefactors and constant terms in order to have the correct limits for 
$k\rightarrow0$, namely the
harmonic oscillator limit.The nature of the trajectories in the $(x-\dot{x})$ 
and $(x-p)$ planes are shown in
Figs.~2. They form closed concentric curves in the region
$|x|<\frac{3\omega}{k}$.

Finally we may point out that because of the form of the second and third
terms in the right hand side of (\ref{lam121}) one may define a modified 
Lagrangian
\begin{subequations}
\begin{align} 
\bar{L}=\frac{1}{k\dot{x}+\frac{k^2}{3}x^2+3\lambda_1},
\qquad\qquad\qquad\qquad\qquad\qquad\qquad\qquad\;\;\qquad\qquad\qquad\qquad\\
\hbox{with the conjugate momentum}
\qquad\qquad\qquad\qquad\qquad
\qquad\qquad\qquad\qquad\qquad\qquad\qquad\qquad\;\nonumber\\
\bar{p}=-\frac{k}{(k\dot{x}+\frac{k^2}{3}x^2+3\lambda_1)^2}
\qquad\qquad\qquad\qquad\qquad\qquad\qquad\;\;\;\qquad\qquad\qquad\qquad\\
\hbox{and the associated Hamiltonian}
\qquad\qquad\qquad\qquad
\qquad\qquad\qquad\qquad\qquad\qquad\qquad\qquad\qquad\nonumber\\
\bar{H}=\frac{2k\dot{x}+\frac{k^2}{3}x^2+3\lambda_1}
{(k\dot{x}+\frac{k^2}{3}x^2+3\lambda_1)^2}
\qquad\qquad\qquad\qquad\qquad\qquad\qquad\quad\;\;\qquad\qquad\qquad\qquad\\
\hbox{or}\qquad\qquad\qquad\qquad\qquad\qquad\qquad\qquad\qquad\qquad
\qquad\qquad\qquad\qquad\qquad\qquad\qquad\qquad\qquad\;\;\;\nonumber\\
H=\bigg(\frac{k}{3}x^2+\frac{3\lambda_1}{k}\bigg)p-2\sqrt{\frac{-p}{k}},
\qquad\qquad\qquad\qquad\qquad\qquad\quad\qquad\qquad\qquad\qquad
\end{align}
\label{lam123a}
\end{subequations}
which again correspond to the equation of motion 
(\ref{lam101}), provided 
$k\neq0$. However in the
limit $k\rightarrow0$, in order to get the correct harmonic oscillator it is
preferable to have the forms (\ref{lam121})-(\ref{lam123}). On the other hand, 
the later
forms (\ref{lam123a}) have the advantage that they give interesting
$\lambda_1=0$ limit, which we discuss in the next section. Consequently, we have
chosen to present both the forms here. 
\section{Unusual features of Eq.(\ref{lam101}) with $\lambda_1>0$}
\label{sec4}
\subsection{Amplitude independent frequency of oscillations}
The remarkable features of the general solution (\ref{lam120}) are that the 
nonlinear system (\ref{lam101}) admits {\it harmonic periodic solutions} 
(\ref{lam120}) and that the {\it frequency of these periodic oscillations}, 
$\nu=\frac{\omega}{2\pi}= \frac{\sqrt{\lambda_1}}{2\pi}$, {\it is completely 
independent of the amplitude or initial condition}, a feature which is generally
quite uncommon for nonlinear oscillators of conservative type.
The form of the solution is shown in Fig.~1. It may be noted that in the
limiting case of vanishing nonlinearity, $k=0$, one recovers the solution
of the linear harmonic oscillator as it should be from Eq.~(\ref{lam120}) and
there is no change in the angular frequency $\omega=\sqrt{\lambda_1}$ even when
$k\neq0$, as can be seen from Eq.~(\ref{lam120}). However, it may be  noted that
in the limit $\lambda_1$ or $\omega\rightarrow0$, $A\rightarrow0$ and so 
$x\rightarrow0$, the trivial solution to (\ref{lam101}) and not to the 
aperiodic bounded solution (\ref{lam102}). Thus at $\lambda_1=0$, a bifurcation 
occurs.

\subsection{The nature of flow function}
Another interesting feature of Eq.~(\ref{lam101}) is that when written as a
system of two first order equations, it takes the form
\begin{subequations}
\begin{align} 
&\dot{x}=y\equiv f_1(x,y), \\
&\dot{y}=-kxy-\frac{k^2}{9}x^3-\lambda_1x\equiv f_2(x,y).
\end{align}
\label{lam124}
\end{subequations}
For $\lambda_1>0$, it has one equilibrium point $(0,0)$ which is of centre type
compatible with the occurrence of nonisolated periodic orbits. Interestingly
the conventional criterion for conservative systems (see for example Refs.
\cite{Lit:1983, Mcc:1993,Dra:1992, Nic:1995}), 
namely, the value of the divergence of the flow function
$\Lambda=\frac{ \partial f_1}{\partial x}+\frac{ \partial f_2}{\partial y}$ for
Eq.~(\ref{lam124}) vanish, fails here as $\Lambda=-kx$. Only the average of
$\Lambda$, namely 
\begin{eqnarray}
\bar{\Lambda}=\frac{1}{T}\int_o^{T}\Lambda dt
=-\frac{k\omega}{2\pi}\int_o^{\frac{2\pi}{\omega}} x dt, \nonumber
\end{eqnarray}
 on actual evaluation using the solution 
(\ref{lam120}), vanishes. Such a generalized criterion that the
average of the flow function vanishes for a conservative system seems to be a
necessary condition as the present example shows.

\section{The aperiodic cases $\lambda_1 \leq 0$}
\label{sec5}
In the previous section, we discussed the dynamics of (\ref{lam101}) for the
case $\lambda_1>0$. In the following we explore the dynamics of the system
(\ref{lam101}) with $\lambda_1 \leq 0$.

For $\lambda_1<0$, both the time-dependent integrals (\ref{lam117}) are
real, from which the explicit solution can be obtained straightforwardly as 
\begin{eqnarray}
x(t)=\bigg(\frac{3\sqrt{|\lambda_1|}(I_1e^{2\sqrt{|\lambda_1|}t}-1)}
{kI_1I_2e^{\sqrt{|\lambda_1|}t}+k{(1+I_1e^{2\sqrt{|\lambda_1|}t})}}\bigg),
\label{lam119a}
\end{eqnarray}
where $I_1$ and $I_2$ are constants. Solution (\ref{lam119a}) clearly 
shows the dissipative/damping/aperiodic nature of the system for 
$\lambda_1\leq0$ in Eq.~(\ref{lam101}).

We now note that system $\lambda_1\leq 0$ still admits the Lagrangian and
Hamiltonian descriptions. In fact the Lagrangian $L$ and the Hamiltonian $H$ 
for the
case $\lambda_1>0$, namely the forms (\ref{lam121}) and (\ref{lam123}), 
respectively, are valid here also. The only difference here is that one has to 
replace $\lambda_1$ with $-|\lambda_1|$ in the respective places in
Eqs.~(\ref{lam121}) and (\ref{lam123}) for the
present case. On the other hand the forms (\ref{lam123a}) are
valid for $\lambda_1>0$ as well as $\lambda_1\leq0$, though not valid in the
limiting harmonic oscillator case $k\rightarrow 0$.

The major difference in the dynamics comes from the nature of the solution. In 
the present case it admits decaying type or aperiodic (front like) solutions 
only, see Fig. 3, where the time of decay or approach to asymptotic value is
independent of the amplitude/initial value, which is once again an unusual
feature for a nonlinear dynamical system.
\begin{figure}[!ht]
\begin{center}
\includegraphics[width=\linewidth]{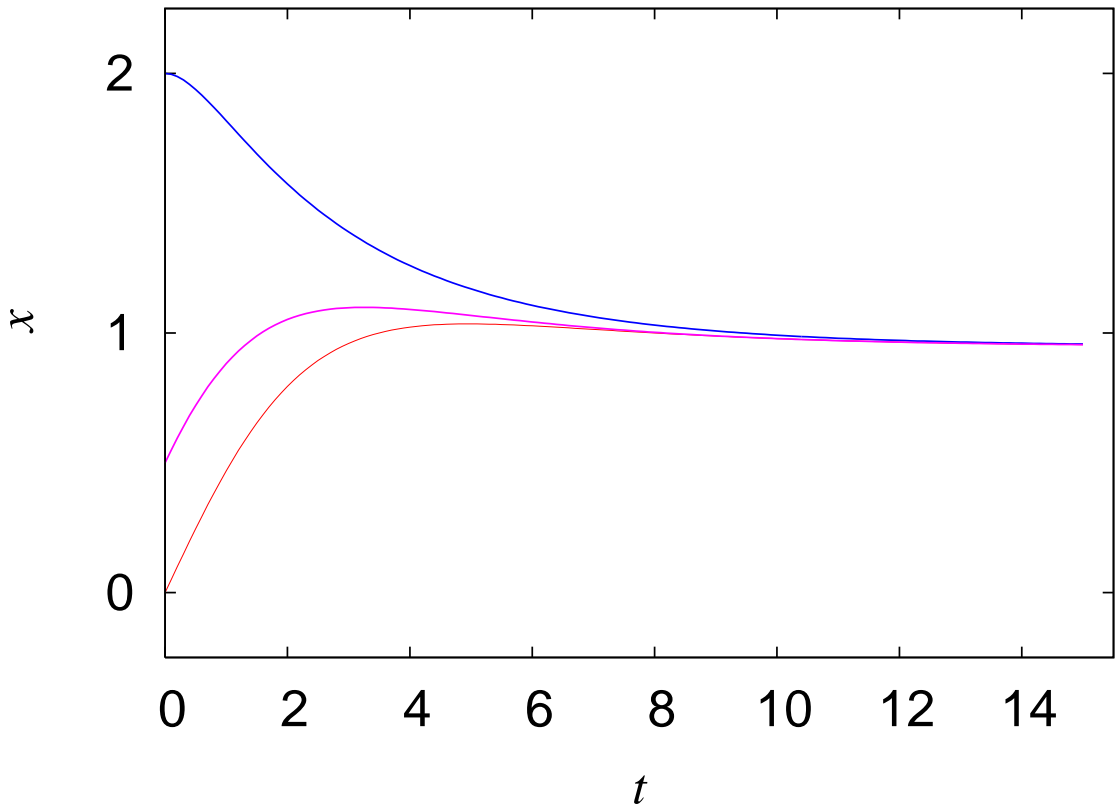}
\caption{(Color online) Solution plot for case $\lambda_1<0$ for different 
initial conditions.}
\end{center}
\end{figure}

The nontrivial general solution for the limiting case $\lambda_1=0$ is given in 
Eq.~(\ref{lam102}).
Interestingly in this case also one can deduce a Lagrangian and the associated
Hamiltonian, which  (from eqs.~(\ref{lam123a})) turns out to be
\begin{equation}
 \hat{L} = \bigg(\frac{1}{k\dot{x}+\frac{k^2}{3}x^2}\bigg),
\end{equation}
and 
\begin{equation}
H = \frac{k}{3}x^2p-2\sqrt{\frac{-p}{k}}, 
\end{equation}
where
\begin{equation}
p = -\frac{k}{(k\dot{x}+\frac{k^2}{3}x^2)^2}.
\end{equation}
The above form of Lagrangian can also be deduced from the form (\ref{lam121}), by
taking the Lagrangian for the limiting case $\lambda_1=0$ as
\begin{equation}
\hat{L}=\lim_{\lambda_1\rightarrow 0}\;\frac{2k^2}{27\lambda_1^3}
(L-\frac{3\lambda_1}{2k}\dot{x}+\frac{9\lambda_1^2}{2k^2}).\nonumber
\end{equation}
The expressions $p$
and $H$ given above can be obtained from this $\hat{L}$.
The solution plot (corresponding to solution (\ref{lam102})) is given in Fig. 4.

One can also calculate the flow function in the case $\lambda_1\leq0$. Again from
Eq. (\ref{lam124}) we find that the divergence of the flow function for
$\lambda_1\leq0$ is $\Lambda=-kx$. But from the general expression for the
solution, one can always choose $x(t)>0$ for all finite $t$. So one might
conclude that the system is dissipative (because $\Lambda<0$) inspite of its 
just proved conservative
nature. In order to overcome this incompatibility, and noting that for the case
$\lambda_1<0$, the dynamical variable $x(t)$ asymptotically tends to a constant value
$x(\infty)=\frac{3\sqrt{\lambda_1}}{k}$, we again define the condition
for conservative flow is that the time average of the divergence of the flow function
with reference to the asymptotic $(t\rightarrow \infty)$ value should vanish:
\begin{eqnarray}
\bar{\Lambda}&=\lim_{T\rightarrow \infty} \frac{1}{T}\int_o^{T}
[\Lambda(t)-\Lambda(\infty)] dt = 0.\label {Lam01}
\end{eqnarray}
In the present cases, we have
\begin{eqnarray}
\bar{\Lambda}&=-\lim_{T\rightarrow \infty} \frac{k}{T}\int_o^{T} 
[x(t)-x(\infty)] dt. 
\label {Lam02}
\end{eqnarray}
For the case $\lambda_1=0$, from the solution (\ref{lam102}) we note that
$x(\infty)=0$ and 
$\int_o^{T}\Lambda(t) dt=-3\log(1+\frac{kT(2I_1+T)}{6I_2})$ 
so that $\bar{\Lambda}=0$ from
(\ref{Lam02}). On the other hand for $\lambda_1<0$, 
$\int_o^{T}\Lambda(t)
dt=-3\sqrt{\lambda_1}T-3\log(\frac{I_1+I_1I_2e^{-T\sqrt{\lambda_1}}
+e^{-2T\sqrt{\lambda_1}}}{(1+I_1+I_1I_2)})$
and $\Lambda(\infty)=-3\sqrt{\lambda_1}$ so that $\bar{\Lambda}$ vanishes here
also.
\begin{figure}[!ht]
\begin{center}
\includegraphics[width=\linewidth]{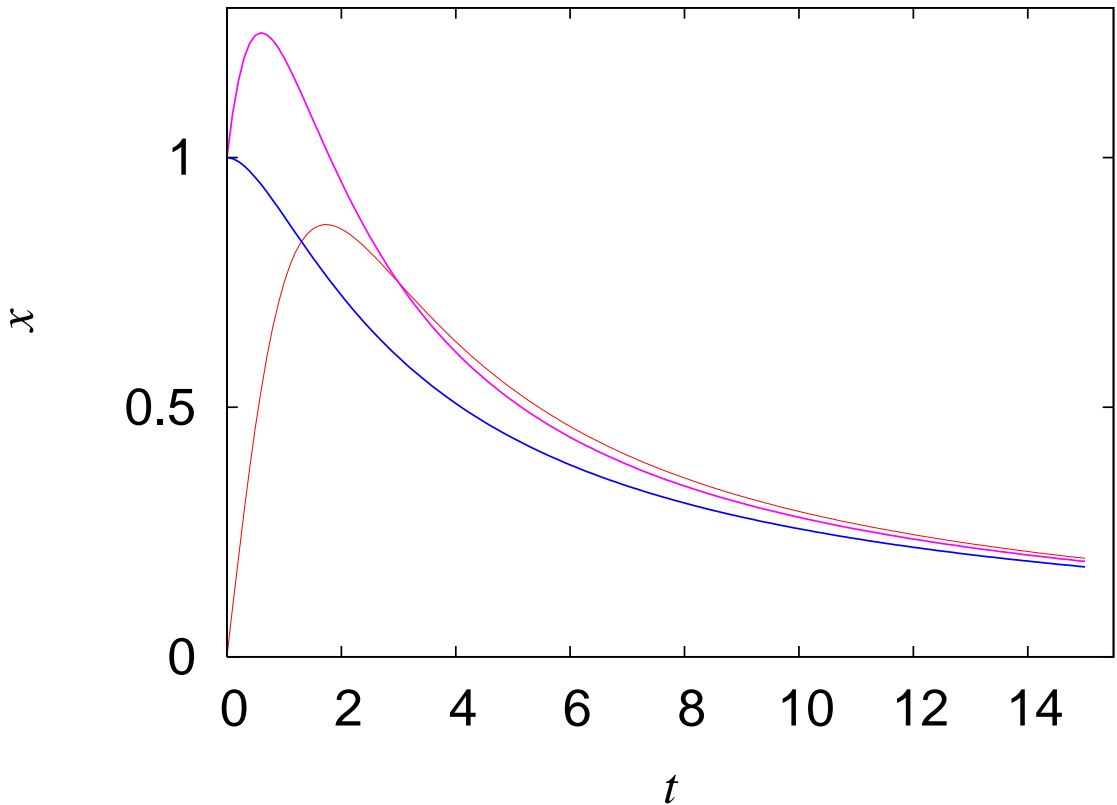}
\caption{ (Color online) Solution plot for case $\lambda_1=0$ for different initial conditions.}
\end{center}
\end{figure}
\section{Linearizing and canonical transformations}
\label{sec6}
At this stage one may also try to trace the reason for the existence of the 
above type 
of remarkable oscillatory behaviour for Eq.~(\ref{lam101}). Under the 
transformation
$x=\frac{3\dot{w}}{kw}$, Eq.~(\ref{lam101}) transforms to a linear third order
differential equation,
\begin{eqnarray}
\dddot{w}+\lambda_1\dot{w}=0, 
\end{eqnarray}
from which also one can trace the solution 
(\ref{lam120}). Further, it gets transformed into a free particle
equation $X''=0$, where prime denotes differentiation with respect to a new
variable $\tau$, with the point transformation 
\begin{eqnarray}
X=(\frac{k}{3\lambda_1}-\frac{1}{\sqrt{-\lambda_1}x})
e^{-\sqrt{-\lambda_1}t},\nonumber\\
\tau=(\frac{k}{3\sqrt{-\lambda_1}}-\frac{1}{x})
e^{\sqrt{-\lambda_1}t}. \qquad \label {lin01}
\end{eqnarray}
 Even then it is very unexpected and unusual to realize 
such nonlinear oscillator systems having simple properties similar to that of 
linear oscillators. 

In addition to the above, one can also linearize the 
equation (\ref{lam101}) through a nonlocal transformation,
\begin{eqnarray}
U=\displaystyle{x e^{\frac{k}{3}\int x(t')dt'}}.
\label {lin02}
\end{eqnarray}
Under this transformation Eq. (\ref{lam101}) gets modifieed to the form of a
linear harmonic oscillator equation,
\begin{eqnarray}
\ddot{U}+\lambda_1U=0.\label {lin03}
\end{eqnarray}
Note that the above transformation (\ref{lin02}) is valid for all the three
cases, namely,
$\lambda_1>0,\;\lambda_1=0$ and $\lambda_1<0$. For $\lambda_1>0$, 
obviously the solution of (\ref{lin03}) is 
\begin{eqnarray}
U=A\sin(\omega t+\delta),\label {lin03a}
\end{eqnarray} 
where $A$ and $\delta$ are arbitrary constants and frequency 
$\omega=\sqrt{\lambda_1}$,
which is independent of the amplitude. Consequently, from the forms of the 
general solutions (\ref{lin03a}) and (\ref{lam120}) we can identify the 
obvious canonical transformation
\begin{eqnarray}
x=\frac{U}{1-\frac{k}{3\lambda_1}P},\label {lin03b}
\end{eqnarray} 
where $P=\dot{U}$, so that using eq.~(\ref{lam122}) for the canonically conjugate
momentum $p$, we have
\begin{eqnarray}
p=P(1-\frac{k}{6\lambda_1}P).\label {lin03c}
\end{eqnarray}
It is straightforward to check that when $U$ and $P$ are canonical so do $x$ and
$p$ (and vice versa) and that the Hamiltonian $H$ in eq.~(\ref{lam123}) can be 
rewritten as the
standard linear harmonic oscillator Hamiltonian
\begin{eqnarray}
H=\frac{1}{2}(P^2+\lambda_1U^2).\label {lin03d}
\end{eqnarray}

More interestingly, even for $\lambda_1<0$, making use of similar argument for
the aperiodic bounded solution (\ref{lam119a}), we can write the canonical
transformation
\begin{eqnarray}
x=\frac{U}{1+\frac{k}{3\lambda_1}P},\qquad \;\;
p=P(1+\frac{k}{6\lambda_1}P),\label {lin03e}
\end{eqnarray}
where $P=\dot{U}$, so that the Hamiltonian (\ref{lam123}) (for $\lambda_1<0$) is
mapped on to the unbounded (wrong sign) linear harmonic oscillator
\begin{eqnarray}
H=\frac{1}{2}(P^2-\omega^2U^2),\;\;\;\omega=\sqrt{|\lambda_1|}.\label {lin03f}
\end{eqnarray}

Finally for the $\lambda_1=0$ case, we can identify the canonical transformation
\begin{eqnarray}
x=\frac{3P}{kU},\qquad\;\;
p=-\frac{kU^2}{6},\label {lin03h}
\end{eqnarray}
so as to transform to a freely falling particle Hamiltonian 
$H=\frac{P^2}{2}-\sqrt{\frac{2}{3}}U$. However, we suspect that there may exist
a canonical transformation for this specific case $\lambda_1=0$, which will take
the corresponding form of Eq.~(\ref{lam101}) to a free particle, though we could
not succeed to find it so far.

Of course, it is well known that 
linearization in general does not ensure
preservation of properties of linear systems in the nonlinear case. The 
classical example is the Cole-Hopf transformation through which the nonlinear 
Burgers equation is transformed into the linear heat equation \cite{Lak:2003}. 
This is also true in the case of various soliton equations such as Korteweg-de
Vries, sine-Gorden, nonlinear Schr$\ddot{o}$dinger and other equations which are all
linearizable in certain sense. Nevertheless in the present example of Eq.
(\ref{lam101}), the amplitude independence of frequency does indeed get
preserved.  These considerations also raise 
the question of identifying the most general class of nonlinear dynamical 
systems of the form
$\ddot{x}=f(x,\dot{x},t)$ that has solutions whose frequency is independent of the
amplitude as in the case of the linear harmonic oscillator, a point which is 
under study presently.

\section{Generalization and other examples}
\label{sec7}
\subsection{Generalization}
We also observed that the above exceptional properties admitted by 
Eq. (\ref{lam101}) is
also common for a class of nonlinear systems. In particular we find that the
modified Emden type equation with linear term and constant external forcing
\begin{eqnarray}            
\ddot{x}+(k_1x+k_2)\dot{x}+\frac {{k_1}^2}{9}x^3+\frac{k_1k_2}{3}x^2
+\lambda_1 x+\lambda_2=0,
\label {lam125}
\end{eqnarray}
where $k_1,k_2,\lambda_1$ and $\lambda_2$ are parameters, for which also one 
can obtain similar propertes as discussed above for 
Eq.~(\ref{lam101}) when the parameters satisfy the specific condition
$\lambda_1=\frac{2k_2^2}{9}+\frac{k_1\lambda_2}{k_2}$.
To be specific, for the 
case $\frac{k_1\lambda_2}{k_2}-\frac{k_2^2}{9}>0$, the explicit periodic 
solution takes the form
\begin{eqnarray}
\bar{x}(t)=\frac{A\sin{(\bar{\omega} t+\delta)}}
{1-(\frac{k}{3\bar{\omega}})A\cos{(\bar{\omega} t+\delta)}},\qquad\qquad\qquad 
\nonumber\\ 
\qquad\qquad 0\leq A <\frac{3\bar{\omega}}{k},\qquad 
\bar{\omega}=\sqrt{\lambda_1-\frac{k_2^2}{3}},
\label{lam120a}
\end{eqnarray} 
where $\bar{x}(t)=x(t)+\frac{k_2}{k_1}$. For 
$\frac{k_1\lambda_2}{k_2}-\frac{k_2^2}{9}<0$, the system becomes aperiodic.

\subsection{Comparison with other nonlinear oscillators}
Finally, it is of interest to compare the dynamics of Eq.~(\ref{lam101}) with
another interesting
nonlinear oscillator \cite{Lak:2003, Mat:1974} of the form
\begin{eqnarray}            
\ddot{x}-\frac{\lambda x\dot{x}^2}{1+\lambda x^2}
+\frac{\omega_0^2x}{1+\lambda x^2}=0,
\label {lam126}
\end{eqnarray}
which possesses exact periodic solution $x(t)=A \sin(\Omega t+\delta)$,
where the frequency $\Omega=\frac{\omega_0}{\sqrt{1+\lambda A^2}}$, exhibiting
the characteristic amplitude-dependent frequency of nonlinear oscillators 
inspite of the sinusoidal nature of the solution Eq.~(\ref{lam126}). 
Equation~(\ref{lam126}) has a natural generalization
in three dimensions \cite{Lak:1976} and these systems can be also quantized 
exhibiting many
interesting features and can be interpreted as an oscillator constrained to 
move on a three-sphere \cite{Hig:1976}. Several generalizations of these 
systems to $N$-degrees of freedom are also possible \cite{Senthil}. But all 
these nice properties are subject to the fact that the
frequency is amplitude-dependent. It is only Eq.~(\ref{lam101}) which admits the
amplitude independence property of frequency for a nonlinear oscillator. Studies
of such systems will have important implications in developing nonlinear systems
whose frequency remain unchanged even with the addition of suitable
nonlinearity.
Also, Eq.~(\ref{lam126}) when written as a system of two first order equations 
of the form (\ref{lam124}), with 
\begin{eqnarray} 
f_1(x,y)=y 
\qquad\nonumber 
\mbox{and} 
\quad\qquad\nonumber 
f_2(x,y)=\frac{\lambda xy^2}{1+\lambda x^2}
-\frac{\omega_0^2x}{1+\lambda x^2}, \nonumber 
\end{eqnarray}
the divergence of the flow function 
$\Lambda=\frac{2\lambda xy}{1+\lambda x^2}\ne0$. On the other hand 
Eq.~(\ref{lam126}) is also a Hamiltonian system \cite{Mat:1974} with the
Hamiltonian given by 
\begin{eqnarray} 
H=\frac{1}{2}[p^2(1+\lambda x^2)
+\frac{\omega_0^2x}{1+\lambda x^2}], \nonumber 
\end{eqnarray}
 eventhough $\Lambda\ne0$. Once again
it is only 
\begin{eqnarray}
\bar{\Lambda}=\frac{1}{T}\int_o^{T}\frac{2\lambda xy}{1+\lambda x^2}
dt=0. \nonumber 
\end{eqnarray}
 This again confirms that for a conservative system it is necessary that
the average of the divergence of the flow function vanish.

\section{Conclusion}
\label{sec8}
In this work, we have identified a nonlinear oscillator which exhibits unusual
oscillatory properties: Purely harmonic oscillation whose frequency is completely 
independent of the amplitude and is the  same as that of the linear harmonic 
oscillator. The conservative Hamiltonian nature of the system has been
established, which also necessitates the generalization of the definition of
conservative systems in terms of the divergence of the flow function.  Further,
even for an apparently dissipative type/aperiodic (front like) type system, the
Hamiltonian structure is preserved and the definition of the conservative system
in terms of the divergence of flow function needs to be generalized. We believe 
identification of such nonlinear systems having the basic property 
of linear systems will have considerable practical application as the effect of
higher harmonics is completely suppressed. Also it is of interest to consider 
the quantum 
mechanical version of the system (\ref{lam123}) and its higher dimensional 
generalizations as well as the effect of external forcing and additional 
damping, which are being pursued. 

We have also shown that Eq.~(\ref{lam101})
admits certain interesting geometrical properties. For example, under the
invertible point transformation (\ref{lin01}), Eq.~(\ref{lam101}) gets transformed to
the free particle equation whereas with the appropriate choice of canonical
transformations one can transform it to a
simple harmonic oscillator (or to an unbounded oscillator or to a freely falling
particle depending on the value of the parameter $\lambda_1$) 
equation. Interestingly, the very same nonlinear
oscillator can also be transformed to a third order linear equation through a
nonlocal transformation, while yet another nonlocal transformation transforms it
to a linear harmonic oscillator. As far as our knowledge goes no such single 
nonlinear
oscillator possesses such large class of transformation properties. We believe
that exploring such nonlinear
equations will be highly rewarding in understanding nonlinear systems and
further work is in progress in this direction.

\begin{acknowledgments}
The work of VKC is supported by Council of Scientific and Industrial Research,
India.  The work of MS and ML forms part of a Department of Science 
and Technology, Government of India, sponsored research project. We thank
Professors Anjan Kundu and  P. E. Hydon for critical reading of the paper and 
Professors V. Balakrishnan and Radha Balakrishnan for discussions.
 
\end{acknowledgments}

\end{document}